\documentclass[aps,prb,twocolumn,amsmath,amssymb,longbibliography]{revtex4}  
\usepackage{bbm}
\usepackage{mathrsfs}
\usepackage{amsmath}
\usepackage{amsfonts}
\usepackage[colorlinks=true,citecolor=blue,anchorcolor=blue]{hyperref}
\usepackage{graphicx,epstopdf}
\usepackage{subfigure}
\usepackage{epsfig}
\usepackage{dcolumn}
\usepackage{bm}
\usepackage{color}
\usepackage{natbib}
\usepackage{amssymb}
\usepackage{xcolor}
\usepackage{braket}

\begin{document}

\title{Axionic surface wave in dynamical axion insulators}

\author{Tongshuai Zhu$^{1}$, Huaiqiang Wang$^{1,\ast}$, Dingyu Xing$^{1,2}$ and Haijun Zhang$^{1,2,\dagger}$}

\affiliation{
$^1$ National Laboratory of Solid State Microstructures, School of Physics, Nanjing University, Nanjing 210093, China\\
$^2$ Collaborative Innovation Center of Advanced Microstructures, Nanjing University, Nanjing 210093, China\\
}

\begin{abstract}
The electromagnetic response of three-dimensional topological insulators can be described by an effective axion action with a quantized axion field. Once both time-reversal and inversion symmetries are broken, for example, in antiferromagnetic topological insulators, the axion field becomes dynamical along with magnetic fluctuations. The dynamical axion field, when coupled to electromagnetic fields, can lead to rich physical phenomena. Here, based on the modified Maxwell's equations, we reveal the existence of an exotic type of polariton excitation, termed a surface axion polariton, which propagates in the interface between a dynamical axion insulator and a dielectric. When doping the dynamical axion insulator to be metallic, the coexistence of axion-photon coupling and plasmon-photon coupling will further generate a mixed surface axion plasmon polariton. We also propose a  Kretschmann-Raether configuration to experimentally detect the surface axion polariton. Our result provides an alternative way to study the axion electrodynamics in condensed matter physics.

\end{abstract}

\email{zhanghj@nju.edu.cn;hqwang@nju.edu.cn}

\maketitle

\begin{figure}[t]
  \centering
  \includegraphics[width=3in]{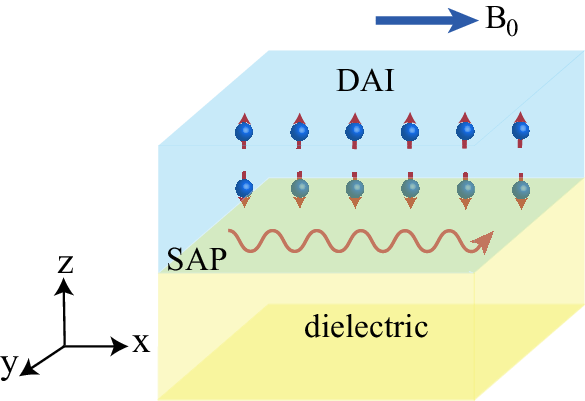}\
  \caption{Schematic illustration of a DAI/dielectric structure. The DAI (dielectric) is located in the $z>0$ ($z<0$) region, with an interface at $z=0$. The red arrows in the DAI represent the magnetic moments. The surface axion polariton (SAP) propagates along the $x$-direction in the interface. }\label{fig1}
\end{figure}

\section{Introduction}  

An axion is a hypothetical particle to solve the charge-parity ($\mathcal{CP}$) problem~\cite{peccei1977cp,weinberg1978new,wilczek1978problem} in quantum chromodynamics. However, the existence of axions in nature has yet to be verified. In condensed matter physics, an effective axion action $S_\theta=\frac{\theta\alpha}{(2\pi)^2}\int d\bm{r}^3dt\bm{E}\cdot\bm{B}$ emerges in three-dimensional (3D) topological insulators, which can be derived from (4+1)-dimensional Chern-Simons theory~\cite{Qi2011topological,Qi2008topological}. Here, $\bm{E}$ and $\bm{B}$ are the electric field and the magnetic field, $\theta$ is the effective axion field, and $\alpha$ is the fine-structure constant. If an insulator preserves the time-reversal symmetry ($\mathcal{T}$) or the inversion symmetry ($\mathcal{P}$), $\theta$ has to be quantized  with the value of $0$ (mod$~ 2\pi)$ for topologically trivial insulators or $\pi$ (mod$~ 2\pi$) for topological insulators. Such a quantized $\theta$ expects to induce various topological magnetoelectric effects~\cite{Qi2008topological,Morimoto2015magnetoelectric,Wang2015QAH,Zirnstein2017magnetoelectric,Rosenberg2010Witten,Coh2011Chern,armitage2019matter}, for example, the quantized magneto-optical Faraday/Kerr rotation~\cite{Maciejko2010topological,Tse2010faraday,Ochiai2012scattering,Karch2009electric,Mal'shukov2013nonlinear,Wu2016faraday,okada2016thrahertz,dziom2017magnetoelectric}, and image magnetic monopole effect~\cite{Qi2009monopole}. Though 3D topological insulators have a quantized axion field $\theta$, the surface states need to be fully gapped to observe those topological magnetoelectric effects. Recently, antiferromagnetic topological insulator MnBi$_2$Te$_4$ was found to be an ideal axion insulator, having both a quantized axion field $\theta$ and a gapped Dirac-cone surface state~\cite{gong2019cpl,otrokov2019nature,zhang2019mbt,li2019sa,liu2020robust,deng2020quantum,chen2019intrinsic,klimovskikh2020tunable,yan2019crystal,hao2019gapless,chen2019topological,zeugner2019chemical,wu2019natural,vidal2019topological,hu2020van,he2020mnbi,lei2020magnetized,rienks2019large}.

When both $\mathcal{T}$ and $\mathcal{P}$ are broken, the axion field would become dynamical with spatial and temporal dependence~\cite{li2010DAF, Wang2013Chiral}. The large dynamical axion field, characterized by a nonzero spin Chern number~\cite{Wang2020heterostructures}, was proposed in antiferromagnetic topological insulators, such as Mn$_2$Bi$_2$Te$_5$~\cite{Zhang2020Mn2Bi2Te5,Cao2o21Growth}, (MnBi$_2$Te$_4$)$_2$/Bi$_2$Te$_3$ superlattice~\cite{Wang2020heterostructures} and MnBi$_2$Te$_4$ films~\cite{zhu2021tunable}. A review of the axion physics in condensed matters can be found in Refs~\cite{Sekine2021axion,nenno2020axion}. A dynamical axion field can give rise to exotic effects, such as the dynamical chiral magnetic effect~\cite{Wilczek1987two,Zhang2020Mn2Bi2Te5,Sekine2016Chiral}, nonlinear level attraction~\cite{Xiao2021cavity}, anomalous magnetoelectric transport in charge-density-wave Weyl semimetals~\cite{gooth2019axionic}, and nonreciprocal surface plasmon polaritons in Weyl semimetals~\cite{Hofmann2016weyl,tsuchikawa2020characterization,Bugaiko2020strained,jalali2019electrodynamics}. Interestingly, in a dynamical axion insulator (DAI), the dynamical axion field can couple linearly to photons, leading to so-called axion polaritons~\cite{li2010DAF} inside the DAI, which will be henceforth termed a bulk axion polariton (BAP). Notably, the BAP spectrum features a tunable forbidden gap, and when the frequency of incident light lies in this gap,  it cannot propagate in the interior of the DAI but expects to be totally reflected~\cite{li2010DAF}. 

Surprisingly, in this paper, we find that a different type of axionic polariton can emerge in the forbidden gap of BAP, with a manifestation of surface waves propagating in the interface between a DAI and a dielectric. This polariton is called a surface axion polariton (SAP) to distinguish it from the BAP. More intriguingly, when the DAI is doped to be metallic, mixed couplings among axions, photons, and plasmons can further generate an exotic surface axion plasmon polariton (SAPP). We have also proposed a feasible experimental setup based on the Kretschmann-Raether configuration~\cite{akimov2017kretschmann,kretschmann1968radiative} to excite and detect the SAP through the minimum reflectivity. 

This paper is organized as follows. In Sec.~\ref{model},  we will give a description of the model consisting of a dynamical axion insulator and a dielectric. In Secs.~\ref{sap} and \ref{sapp},  we will investigate SAP and SAPP, respectively. In Sec.~\ref{experiment}, we will propose a feasible experimental setup to detected the surface axion polariton. We conclude in Sec.~\ref{conclusion}.
\section{Model descriptions}  
\label{model}
To demonstrate the emergence of axionic surface wave, we consider a system where a DAI hosting dynamical axion field is in close contact with a dielectric, as schematically shown in Fig.~\ref{fig1}. The DAI and the dielectric are placed in the $z>0$ and  $z<0$  regions, respectively, with an interface between them located at $z=0$. Due to the breaking of $\mathcal{T}$ and $\mathcal{P}$, the  axion field $\theta$ in the DAI is generically nonquantized, which can be decomposed into a static part and a dynamical part as $\theta(\bm{r},t)=\theta_0+\delta\theta(\bm{r},t)$. Previous works~\cite{li2010DAF, Zhang2020Mn2Bi2Te5,Wang2020heterostructures} have already shown that the spin-wave excitation in DAIs can induce fluctuations of the axion field, since $\delta\theta=\delta m_5/g$, where $\delta m_5 $ is proportional to the amplitude fluctuation of the antiferromagnetic order along the $z$ direction, and $1/g$ is a tunable coefficient defined as $\partial\theta/\partial m_5$. 

In addition, we apply a static external magnetic field along the $x$ direction, given by $\bm{B}_0=B_0\hat{x}$,  in order to generate a linear coupling between the axion field and $x$ component of the electric field $E_x$ of electromagnetic waves~\cite{li2010DAF}.  The electrodynamics of such an axion-photon coupled system can be obtained by the Euler-Lagrangian equation, leading to modified Maxwell's equations and the equation of motion for the axion~\cite{li2010DAF}. For  better clarity, we take the long-wavelength approximation and ignore the dispersion of the axion (the inclusion of axion dispersion will not affect our results qualitatively),  so the linearized equation of motion can be simplified as [see the Supplemental Material (SM)~\cite{SM} for more details]
\begin{equation}
\label{linearizedMaxwell}
\begin{split}
\nabla\times\bm{H}-\frac{1}{c}\frac{\partial\bm{D}}{\partial t}&=\frac{\alpha}{c\pi}(\partial_t\delta\theta)\bm{B}_0\\
\frac{\partial^2}{\partial t^2}\delta\theta+m^2\delta\theta+\Gamma\frac{\partial}{\partial t}\delta\theta&=\Lambda{E_x}{B}_0
\end{split}
\end{equation}
where $\Lambda=\alpha/(8\pi^2g^2J)$, and $J$,  $m$, and $\Gamma$ are the material-dependent stiffness, mass, and damping of the axion mode. The constituent equations are $\bm{D}=\epsilon \bm{E}$ and $\bm{B}=\mu\bm{H}$, where $\epsilon$ and $\mu$ are the dielectric constant and magnetic permeability, respectively. It should be mentioned that the coefficient $1/g$ can be tuned by the strength of the spin-orbit coupling which can be realized by element substitution, and $1/g$ reaches its maximum value near the topological transition point~\cite{Zhang2020Mn2Bi2Te5,Xiao2021cavity}.

\section{Surface axion polariton}
\label{sap}
\begin{figure}[t]
  \centering
  \includegraphics[width=3.4in]{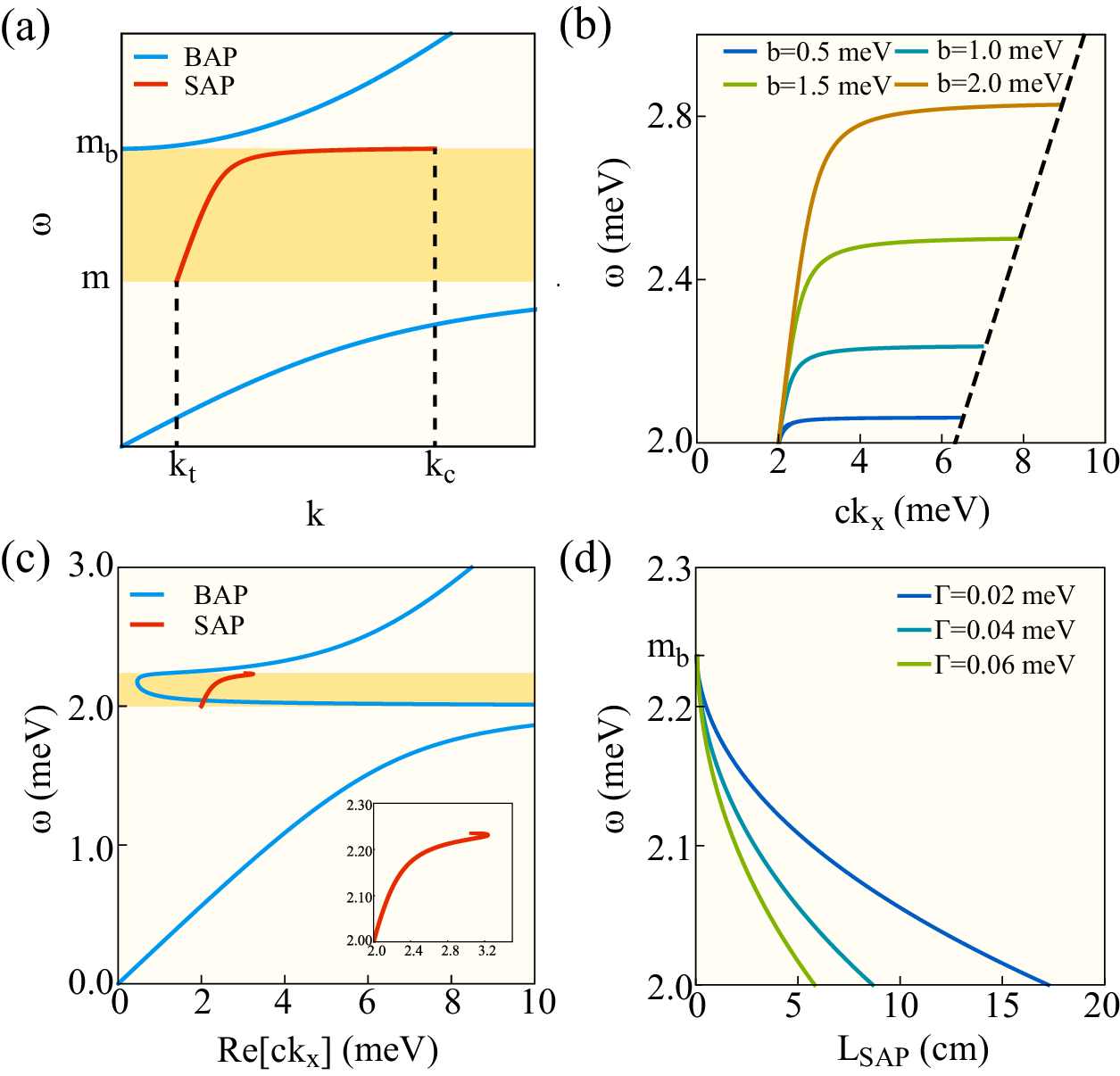}\
  \caption{Spectra of surface axion polariton (SAP) with and without axion damping. (a) SAP spectrum (red lines) in the forbidden gap (shaded in orange) of the bulk axion polariton (BAP) spectrum (blue lines). (b) SAP spectra for different values of axion-photon coupling strengths $b$, where the black dashed line represents the cutoff wave vector. (c) Spectra of BAP (blue lines) and SAP (red lines) with axion damping $\Gamma=0.02$ meV.  Inset: Enlarged view of the SAP spectrum. (d) The propagation length of SAP for different values of axion damping. The parameters in the numerical calculations are chosen as $\epsilon_{\mbox{\tiny I}}=10$, $\epsilon_{\mbox{\tiny II}}=1$, $\mu_{\mbox{\tiny I}}=\mu_{\mbox{\tiny II}}=1$, and $m=2$ meV.}\label{fig2}
\end{figure}

 Before the discussion of SAP, we first recall the BAP for later reference. The BAP originates from the linear coupling between the photon and the axion excitation inside a DAI~\cite{li2010DAF}. Based on Eq.~(\ref{linearizedMaxwell}), the dispersion relation of the BAP with axion damping is given as~\cite{li2010DAF,Xiao2021cavity}
\begin{equation}
 \begin{split}
 k=\sqrt{\frac{\epsilon\mu\omega^2(-b^2-m^2+i\Gamma\omega+\omega^2)}{-m^2+i\Gamma\omega+\omega^2}}
\end{split}
 \end{equation} 
where $k$ and $\omega$ are the wave vector and frequency, respectively, and $b\equiv\alpha B_0/\sqrt{8\pi^3g^2J\epsilon}$ is related to the axion-photon coupling strength, which depends on $B_0$, ${1}/{g}$, and $J$. For typical DAIs such as the MnBi$_2$Te$_4$/Bi$_2$Te$_3$ heterostructure~\cite{Wang2020heterostructures}, $b$ can be tuned in the approximate range of ($0-5$) meV~\cite{Xiao2021cavity} under characteristic values of $\epsilon$ and $B_0$. The dispersion relations of the BAP with and without axion damping are plotted as blue lines in Figs.~\ref{fig2}(a) and \ref{fig2}(c), respectively. The most prominent feature of the BAP dispersion relation is the emergence of a forbidden gap separating two BAP branches with limiting frequencies of $m$ and $m_b\equiv\sqrt{m^2+b^2}$,  shaded in orange in Figs.~\ref{fig2}(a) and ~\ref{fig2}(c). Interestingly, when the frequency of the incident light lies within the forbidden gap of the BAP, total reflection is expected to occur, leading to significantly enhanced reflectivity.

Apart from the BAP propagating in the bulk of the DAI, now we will show that a surface wave is allowed to propagate along the interface between the DAI and dielectric, which amounts to finding a bound state solution to the modified Maxwell's equations under appropriate boundary conditions. Surface waves are strongly localized at the interface and exponentially decaying away from the interface. There are many types of surface waves such as surface plasmon polaritons\cite{barnes2003surface}, surface photon polaritons\cite{LeGall1997Experimental,shen2009surface,Huber2005phonon,li2016reversible}, surface exciton polaritons\cite{Lagois1976exciton}, Dyakonov surface waves\cite{Dyakonov1988new}, and so on \cite{Gollub2005magnetic, camley1987nonreciprocal, yeh1978optical, tomlinson1980surface, takayama2017photonic, bharadwaj2022pico}. We consider the transverse magnetic (TM) mode case at first, where the magnetic field is transverse to the direction of surface wave propagation. The propagation direction of the surface wave is fixed along the $x$-direction. By taking the plane wave ansatz of both the electromagnetic wave and the axion field, the magnetic field at the interface can be written as $H_{y}^\eta=H_0^\eta e^{i k_xx-i\omega t}e^{s k^\eta_z z}$, where $\eta=$ I, II denote the $z>0$ and $z<0$ regions, respectively, $k_z^\eta$ is the decay constant, and $s=-1(+1)$ for $\eta=$ I, II.  By solving the modified Maxwell's equations with the boundary conditions of $H_y^{\mbox{\tiny I}}=H_y^{\mbox{\tiny II}}$ and $E_x^{\mbox{\tiny I}}=E_x^{\mbox{\tiny II}}$, the ratio between the two decay constants is obtained as~\cite{SM} 
\begin{equation}
\begin{split}
\frac{k_z^{\mbox{\tiny I}}}{k_z^{\mbox{\tiny II}}}=-\frac{\epsilon_{\mbox{\tiny I}}^{\mbox{\tiny eff}}}{\epsilon_{\mbox{\tiny II}}},
\end{split}
\end{equation}
where $\epsilon_{\mbox{\tiny I}}^{\mbox{\tiny eff}}=\epsilon_{\mbox{\tiny I}}(1-\frac{b^2}{\omega^2-m^2+i\Gamma\omega})$. To get a stable surface wave solution localized at the $z=0$ interface, both $k_z^{\mbox{\tiny I}}$ and $k_z^{\mbox{\tiny II}}$ must be positive, thus requiring opposite signs between  $\epsilon_{\mbox{\tiny I}}^{\mbox{\tiny eff}}$ and $\epsilon_{\mbox{\tiny II}}$.  Since the dielectric constant $\epsilon_{\mbox{\tiny II}}$ in the dielectric is always positive, the effective dielectric constant $\epsilon_{\mbox{\tiny I}}^{\mbox{\tiny eff}}$ needs to be negative. This can only be satisfied when $m<\omega<m_b$, i.e., the frequency of the surface wave must exist in the forbidden gap of BAP. 

The surface wave excitation results from the axion-photon coupling and exists only at the interface. Therefore it is termed SAP to distinguish it from BAP. The dispersion relation of SAP can be obtained as~\cite{SM} 
\begin{equation}
\label{SAPdispersion}
\begin{split}
k_x=\omega\sqrt{\frac{\epsilon_{\mbox{\tiny I}}\mu_{\mbox{\tiny I}}\epsilon_{{\mbox{\tiny II}}}^2-\epsilon_{{\mbox{\tiny I}}}\epsilon_{\mbox{\tiny II}}\epsilon_{\mbox{\tiny I}}^{\mbox{\tiny eff}}\mu_{\mbox{\tiny II}}}{\epsilon_{\mbox{\tiny II}}^2-\epsilon_{\mbox{\tiny I}}\epsilon_{\mbox{\tiny I}}^{\mbox{\tiny eff}}}},
\end{split}
\end{equation}
where $\mu_{\mbox{\tiny I}}$ and $\mu_{\mbox{\tiny II}}$ are the permeabilities of the DAI and the dielectric, respectively. Since $\omega$ is constrained in the range between $m$ and $m_b$ of the forbidden gap of BAP, there exists a threshold wave vector of $k_t=m\sqrt{\epsilon_{\mbox{\tiny II}}\mu_{\mbox{\tiny II}}}$ when $\omega\rightarrow m$ and a cutoff wave vector of $k_c=\sqrt{\epsilon_{\mbox{\tiny I}}\mu_{\mbox{\tiny I}}(b^2+m^2)}$ when $\omega\rightarrow m_b$, as schematically shown by the red lines in Fig.~\ref{fig2}(a). This is different from the dispersion relation of a surface plasmon polariton starting at zero wave vector and approaching an asymptotic frequency at large wave vectors. In Fig.~\ref{fig2}(b), we present the dispersion relation of SAP for different values of $b$, where the $b$-dependent cutoff wave vector has been explicitly plotted as the black dashed line. It can be seen that a larger $b$ leads to a larger range of allowed SAP frequency as well as a larger cutoff wave vector. Consequently, topological DAIs with a large dynamical axion field and a large $b$ are highly preferred candidates for SAP.

When taking the axion damping into account, the propagation vector becomes complex as $k_x=k_x'+ik_x''$. The dispersion relation of SAP with a typical value of axion damping $\Gamma=0.02$ meV is shown by the red line in Fig.~\ref{fig2}(c). Notably, in contrast to the undamped case, the cutoff wave vector of SAP is no longer at $\omega_b$ [see the inset of Fig.~\ref{fig2}(c)]. In addition, the imaginary part of $k_x$ will result in a finite propagation length of the SAP, which is defined as $L_{\mathrm{SAP}}\equiv1/(2k_x'')$. In Fig.~\ref{fig2}(d), we choose different values of $\Gamma$ to plot the propagation length $L_{\textrm{SAP}}$ as a function of $\omega$ over the whole frequency range of SAP, where $L_{\textrm{SAP}}$ lies in the cm ranges and  increases with decreasing $\omega$.

\begin{figure}[t]
  \centering
  \includegraphics[width=3.4in]{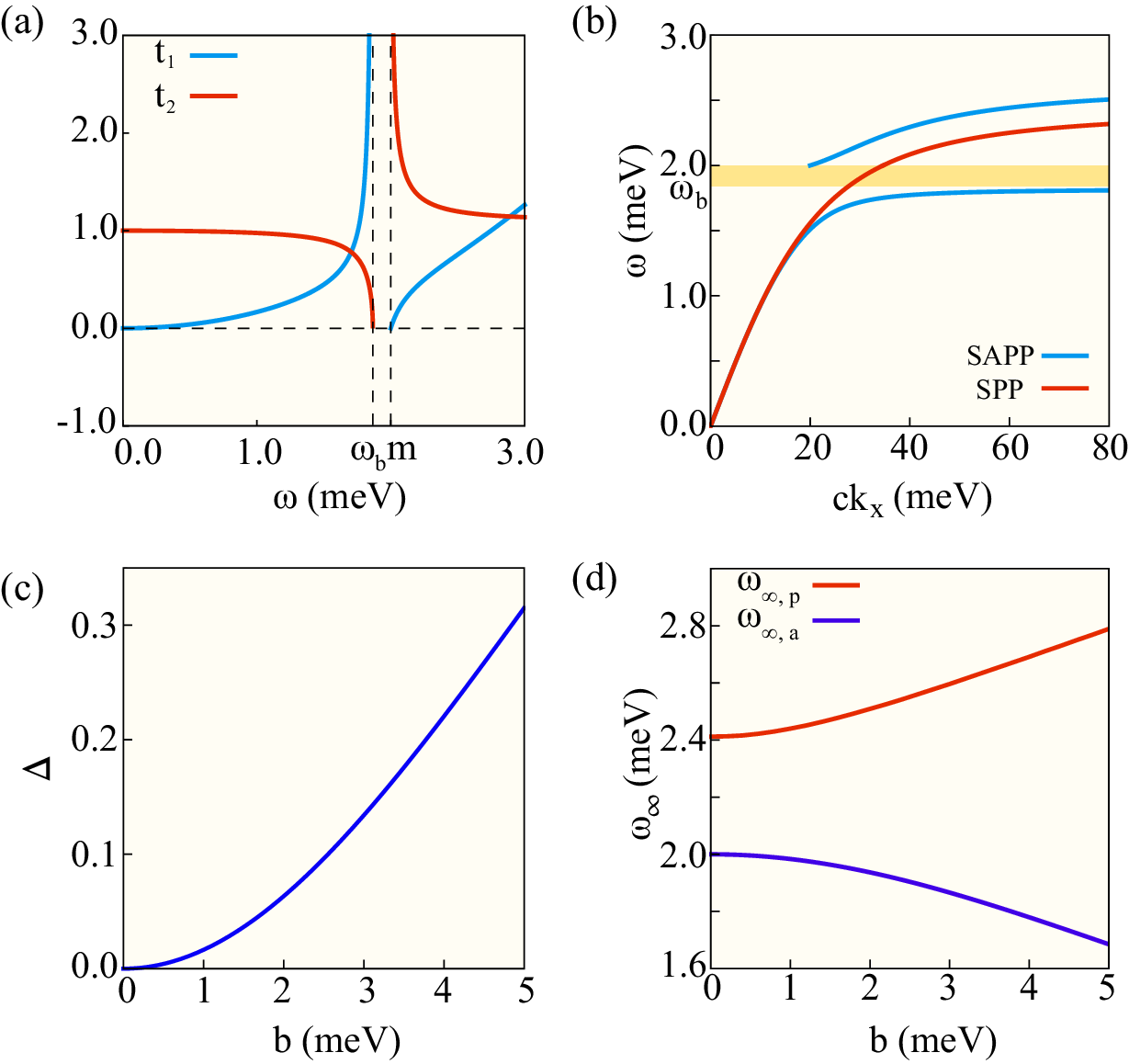}\
  \caption{ Spectra of the surface axion plasmon polariton (SAPP). (a) The two ratios, $t_1$ (blue) and $t_2$ (red), vs frequency with $b=3$ meV. Both $t_1$ and $t_2$ exhibit a gap between $\omega_b$ and $m$, where they become purely imaginary. (b) Spectrum of the SAPP without axion-photon coupling ($b=0$ meV, red line), which is reduced to that of SPP.  Spectrum of SAPP with axion-photon coupling ($b=3$ meV, blue lines). The orange area represents the forbidden gap of SAPP. (c) The magnitude of the forbidden gap $\Delta$ of SAPP as a function of $b$. (d) The asymptotic frequencies of $\omega_{\infty,p}$ (green) and $\omega_{\infty,a}$ (blue) with varying $b$. The parameters are $\epsilon_{\mbox{\tiny I}}=10$, $\epsilon_{\mbox{\tiny II}}=100$, $\mu_{\mbox{\tiny I}}=\mu_{\mbox{\tiny II}}=1$, $m=2$ meV, and $\omega_p=8$ meV. }\label{fig3}
\end{figure}

\begin{figure*}[htbp]
   \centering
   \includegraphics[width=6.2in]{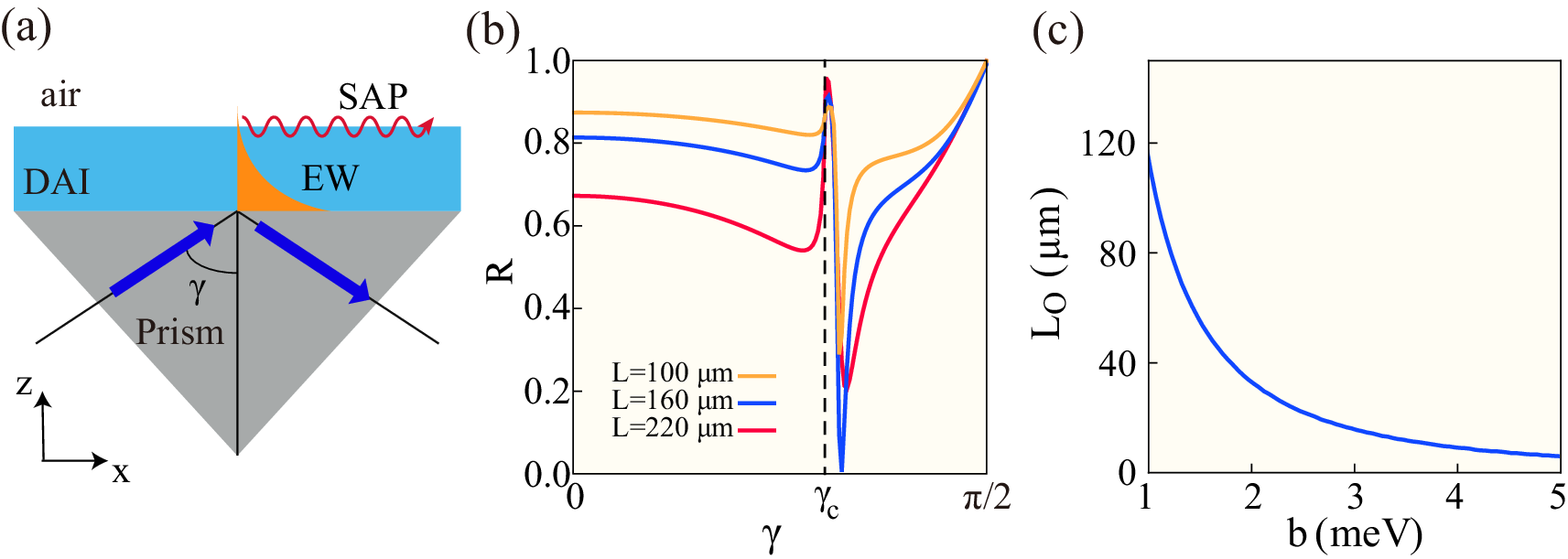}\
   \caption{Excitation and detection of a surface axion polariton (SAP). (a) Experimental setup of the Kretschmann-Raether configuration, where a dynamical axion insulator (DAI) is placed between air and a prism. Light is incident onto the prism with an angle of $\gamma$, which transforms into an evanescent wave (EW) when $\gamma$ exceeds the critical angle of total internal reflection. The SAP is excited in the interface between the DAI and air, which propagates along the $x$-direction. (b) The reflectivity of the incident light with a frequency of $\omega=2.1$ meV as a function of the incident angle for different thicknesses $L$ of the DAI. The dashed black line marks the critical angle. Resonant excitation of SAP manifests as the minimum reflectivity when $L=160 \mu$m. (c) The optimal thickness of the DAI exhibits the minimum reflectivity as a function of $b$.}\label{fig4}
 \end{figure*}

\section{Discussion} 
\label{sapp}
More interestingly, as pointed out in Refs.~\cite{Karch2011plasmons,Robert2013surface}, under modified boundary conditions induced by the nontrivial axion term, a boundary charge term appears at the boundary of the DAI, which can lead to a mixing of TM and transverse electric (TE) modes. In this case, the plane wave ansatz for the TM and TE modes can be written as
\begin{equation}
\begin{split}
H_y^\eta=H_0^\eta e^{i k_xx-i\omega t}e^{s k^{\eta}_{z,\mbox{\tiny TM}} z},\\
E_y^\eta=E_0^\eta e^{i k_xx-i\omega t}e^{s k^{\eta}_{z,\mbox{\tiny TE}} z},
\end{split}
\end{equation} 
where $\eta=$ I, II denote the $z>0$ and $z<0$ regions, respectively,  $s=-1 \ (+1) $ for $\eta=$ I (II), and $k^{\eta}_{z,\mbox{\tiny TM}}$ ($k^{\eta}_{z,\mbox{\tiny TE}}$) are the decay constants for the TM (TE) mode. For convenience, we define two dimensionless ratios between the decay constants as $t_1=k_{z,\mbox{\tiny TM}}^{\mathrm{{\mbox{\tiny I}}}}/k^{\mathrm{{\mbox{\tiny I}}}}_{z,\mbox{\tiny TE}}$, $t_2=k^{\mathrm{{\mbox{\tiny II}}}}_{z,\mbox{\tiny TM}}/k^{\mathrm{{\mbox{\tiny I}}}}_{z,\mbox{\tiny TM}}$. The detailed expressions of $t_1$  and $t_2$ can be obtained through the modified boundary conditions of $H^{\mbox{\tiny I}}_{x,y}-\kappa\theta_{\mbox{\tiny I}} E^{\mbox{\tiny I}}_{x,y}=H^{\mbox{\tiny II}}_{x,y}-\kappa\theta_{\mbox{\tiny II}} E^{\mbox{\tiny II}}_{x,y}$~\cite{SM}.  To find a surface wave solution where all the electromagnetic modes are localized at the interface, $t_1$ and $t_2$ must be positive. However, since the dielectric constant $\epsilon_{\mbox{\tiny I}}$ of DAI is positive, such a condition cannot be satisfied~\cite{SM}, thus forbidding the existence of surface waves in this case. 
 
Nevertheless, the above issue can be circumvented by doping the DAI to shift its Fermi level to the conduction or valence bands to be metallic. Its dielectric function can then be described by the Drude model with frequency-dependent $\epsilon_{{\mbox{\tiny D}}}=\epsilon_{\mbox{\tiny I}}(1-\frac{\omega_p^2}{\omega^2+i\gamma\omega})$, where $\omega_p^2=\frac{ne^2}{\epsilon_0\epsilon_{\mbox{\tiny I}}m^*}$ is the plasma frequency of the charge carriers. Here, $m^*=m_e$ is the effective mass, $\epsilon_0$ is the vacuum permittivity, and $n$ is the carrier concentration. Similar to a typical metal case, the Drude dielectric function becomes negative in certain frequency ranges below $\omega_p$, within which the condition of positive $t_1$ and $t_2$ can be satisfied~\cite{SM}. The surface wave excitation found in this case is called as SAPP, due to mixed couplings among axions, photons and surface plasmons. The dispersion relation of SAPP is obtained as~\cite{SM}
\begin{equation}
k_x=\omega\sqrt{\frac{t_2^2\epsilon_{{\mbox{\tiny D}}}^{{\mbox{\tiny eff}}}\epsilon_{\mbox{\tiny D}}\mu_{\mbox{\tiny I}}-\epsilon_{\mbox{\tiny II}}\mu_{\mbox{\tiny II}}\epsilon_{\mbox{\tiny D}}}{t_2^2\epsilon_{{\mbox{\tiny D}}}^{{\mbox{\tiny eff}}}-\epsilon_{\mbox{\tiny D}}}},
\end{equation}
where $\epsilon_{{\mbox{\tiny D}}}^{{\mbox{\tiny eff}}}=\epsilon_{{\mbox{\tiny D}}}[(1-\frac{\epsilon_{\mbox{\tiny I}}b^2}{\epsilon_{{\mbox{\tiny D}}}(\omega^2-m^2)})]$.
Interestingly, as shown in Fig.~\ref{fig3}(a), it is found that $t_1$ (red lines) and $t_2$ (blue lines) become purely imaginary when the frequency $\omega$ satisfies $\omega_b<\omega<m$ (the detailed expression of the $b$-dependent $\omega_b$ is given in the SM~\cite{SM}). This will result in a forbidden gap $\Delta=m-\omega_b$ in the SAPP spectrum, as shaded in orange in Fig.~\ref{fig3}(b). 
The magnitude of the forbidden gap $\Delta$ increases with increasing $b$, as presented in Fig.~\ref{fig3}(c). It should be pointed out that in the absence of axion-photon coupling ($b=0$ meV), the SAPP will be reduced to a traditional surface plasmon polariton without the forbidden gap, as reflected by the red line in Fig.~\ref{fig3}(b). In addition, there are two limiting frequencies for a sufficiently large wave vector $k_x$. One is the axionic frequency $\omega_{\infty,a}$ and the other is the plasmonic frequency $\omega_{\infty,p}$.  Figure~\ref{fig3}(d) shows the two asymptotic frequencies as a function of $b$, where it can be seen that the plasmonic (axionic) frequency $\omega_{\infty,p}$ ($\omega_{\infty,a}$) increases (decreases) with increasing $b$.

\section{Experimental Detection}
\label{experiment}
 To experimentally excite SAP by light, both frequency and wave vector must be matched. However, since the SAP wave vector is always larger than that of light in free space, an increment of light wave vector is required. This can be achieved by the experimental setup based on the classical Kretschmann-Raether configuration~\cite{akimov2017kretschmann,kretschmann1968radiative}, as illustrated in Fig.~\ref{fig4}(a), where the DAI is placed between a prism and air. When the angle of the incident light onto the prism is larger than the critical angle $\gamma_c$ of total internal reflection, the incident light can be converted to evanescent wave  to increase the wave vector. Resonant excitation of SAP in the DAI/air interface can be realized by tuning the incident angle to match the parallel wave vector, which will manifest as a minimum in reflectivity. In Fig.~\ref{fig4}(b), we plot the reflectivity of the light as a function of the incident angle under different thicknesses $L$ of the DAI with fixed $b=1$ meV. The minimum reflectivity shows up at $L=160\ \mu$m.  Further, in Fig.~\ref{fig4}(c), we present the  optimal thickness of the minimum reflection with varying $b$, which is found to decrease with increasing $b$. Since $b$ can be tuned by an external magnetic field or by internal element substitution, this will lead to a tunable  optimal thickness for resonant excitation, which might be a smoking gun to detect the SAP.

\section{Conclusion} 
\label{conclusion}
In summary, we have proposed the SAP as another type of polariton propagating in the surface of a DAI, which lies in the forbidden gap of the BAP. When doping the DAI to the metallic phase, the simultaneous presence of axion-photon coupling and plasmon-photon coupling leads to the SAPP as a mixed type of polariton. An experimental scheme of the Kretschmann-Raether configuration was proposed to experimentally detect the SAP, where resonant excitation of SAP shows up as a minimum reflectivity. The proposal of SAP could further facilitate the study of axion electrodynamics in condensed matter physics. 

\section*{ACKNOWLEDGEMENTS}
This work is supported by National Key Projects for Research and Development of China (Grant No.2021YFA1400400), the Fundamental Research Funds for the Central Universities (Grant No. 020414380185), Natural Science Foundation of Jiangsu Province (No. BK20200007), the Natural Science Foundation of China (No. 12074181, No. 11834006, and No. 12104217) and the Fok Ying-Tong Education Foundation of China (Grant No. 161006).

T.~Z. and H.~W. contributed equally to this work.

\bibliography{ref}

\end{document}